\documentclass[twocolumn,twocolappendix]{aastex7}
\usepackage{graphicx}
\usepackage{amsfonts,amssymb,amsmath}
\usepackage{xcolor}

\begin{document}

\title{Eccentricity as a signature of hierarchical subsolar-mass mergers in collapsar disks}

\author[0000-0003-3829-967X]{Jiaxi Wu}
\email[show]{jiaxiwu@caltech.edu}
\affiliation{TAPIR, Mailcode 350-17, California Institute of Technology, Pasadena, CA 91125, USA}
\author[0000-0002-0491-1210]{Elias R. Most}
\email[show]{emost@caltech.edu}
\affiliation{TAPIR, Mailcode 350-17, California Institute of Technology, Pasadena, CA 91125, USA}
\affiliation{Walter Burke Institute for Theoretical Physics, California Institute of Technology, Pasadena, CA 91125, USA}
\author[0000-0002-5767-3949]{Nils L. Vu}
\email{nilsvu@caltech.edu}
\affiliation{TAPIR, Mailcode 350-17, California Institute of Technology, Pasadena, CA 91125, USA}
\affiliation{Walter Burke Institute for Theoretical Physics, California Institute of Technology, Pasadena, CA 91125, USA}
\author[0000-0003-4557-4115]{Nils Deppe} 
\email{}
\affiliation{Laboratory for Elementary Particle Physics, Cornell University, Ithaca, New York 14853, USA}
\affiliation{Department of Physics, Cornell University, Ithaca, NY, 14853, USA}
\affiliation{Cornell Center for Astrophysics and Planetary Science, Cornell University, Ithaca, New York 14853, USA}
\author[0000-0001-5392-7342]{Lawrence E. Kidder}
\email{}
\affiliation{Cornell Center for Astrophysics and Planetary Science, Cornell University, Ithaca, New York 14853, USA}
\author[0000-0003-2426-8768]{Kyle C. Nelli} 
\email{}
\affiliation{TAPIR, Mailcode 350-17, California Institute of Technology, Pasadena, CA 91125, USA}
\author[0000-0001-5059-4378]{William Throwe} 
\email{}
\affiliation{Cornell Center for Astrophysics and Planetary Science, Cornell University, Ithaca, New York 14853, USA}

\begin{abstract}
In this work, we investigate gravitational-wave signatures of a proposed subsolar-mass merger scenario resulting from fragmentation inside a collapsar accretion disk. This scenario has gained recent interest with the electromagnetic transient AT2025ulz, a possible superkilonova counterpart candidate to the sub-threshold gravitational wave event S250818k. One prediction of fragmentation is the formation of multiple smaller neutron-star fragments, some of which might merge hierarchically. Such mergers are expected not only to produce individual electromagnetic counterparts, but also, because of their repeated capture and merger dynamics, to impart kicks to the system and thereby drive orbital eccentricity. By performing numerical relativity simulations of hierarchical subsolar-mass compact-object mergers modeled as black holes in a disk-like geometry consistent with this scenario, we demonstrate the build-up of potentially large eccentricity for the final merger, of order $e \simeq 0.6$ initially, and show that, because of the short lifetime of the system, a substantial part of this eccentricity , up to $e\simeq 0.1$, can survive until  the final neutron star -- black hole merger in the general case. As a result, future detections of eccentricities in potential subsolar-mass gravitational-wave candidate events would be a strong indicator for a hierarchical formation scenario. In the extreme case, where we observe repeated mergers to lead to the formation of a solar-mass neutron star, the expected binary parameters can be in a regime similar to those of the eccentric neutron star -- black hole merger event GW200105.
\end{abstract}
\section{Introduction}

Mergers of neutron stars with other compact objects are exciting sources for multi-messenger astronomy \citep{Lattimer:1974slx, Eichler:1989ve}.
So far, the detections of the neutron-star merger events GW170817 and GW190425 have highlighted the exciting prospect posed by comparable-mass neutron stars \citep{LIGOScientific:2017vwq, LIGOScientific:2020aai}. GW170817 in particular was accompanied by a kilonova \citep{Cowperthwaite:2017dyu, Chornock:2017sdf, Kasliwal:2017ngb, Pian:2017gtc, Drout:2017ijr}, a short-duration gamma-ray burst \citep{LIGOScientific:2017zic, Savchenko:2017ffs}, a radio counterpart \citep{Hallinan:2017woc, Alexander:2017aly, Mooley:2017enz}, and an X-ray afterglow \citep{Troja:2017nqp, Margutti:2017cjl}. While neutron stars merging with stellar-mass black holes can have similar prospects if the black hole is rapidly spinning or less massive than about $5 M_\odot$, most discovered events likely fall outside of this regime \citep{LIGOScientific:2021qlt, Martineau:2024zur}, and have not been observed to be accompanied by counterparts \citep{Andreoni:2019qgh, Anand:2020eyg, Ahumada:2025ubp}.
While these likely represent a large fraction of the expected mass spectrum, some merger regimes remain poorly understood. On the upper-mass end, neutron stars are bounded by a maximum mass, subject to constraints from the properties of dense nuclear matter \citep{Lattimer:2015nhk, MUSES:2023hyz}, causality \citep{Rhoades:1974fn}, pulsar timing \citep{NANOGrav:2019jur, Demorest:2010bx}, and potentially also neutron-star merger observations \citep{Margalit:2017dij, Rezzolla:2011da, Ruiz:2016rai, Shibata:2019ctb}. From the standpoint of stellar evolution, the possible existence of a lower mass gap would support this picture \citep{Fryer:1999mi, Fryer:1999ht, Farr:2010tu, Kreidberg:2012ud, Zevin:2020gma}, although recent gravitational-wave observations of compact objects in this range have challenged our understanding \citep{LIGOScientific:2020zkf, LIGOScientific:2021qlt}; see also \citet{Burrows:2023nlq} for advances in core-collapse supernova modeling. 
At the other end, neutron stars are expected to have a lower mass bound, coming both from stellar evolution \citep{Suwa:2018uni, Muller:2024aod} and from the fact that a minimal neutron degeneracy needs to be maintained for the star to be stable against runaway $\beta-$decay of neutrons and ultimate explosion \citep{1989ApJ...339..318C, 1991ApJ...369..422C}.
It is therefore generally difficult to produce subsolar-mass neutron stars in most astrophysical contexts. One potential scenario involves the formation of a collapsar accretion disk, the outskirts of which can become gravitationally unstable and form neutron-star fragments \citep{Piro:2006ja, Metzger:2024ujc}, akin to the gravitational-instability channel of giant planet formation in protoplanetary disks \citep{Boss:1997}.
While the details of this channel are highly sensitive to rapid neutronization of the fragment, numerical simulations have lent credibility to the basic mechanism \citep{Chen:2025uwd}.
In fact, it is then predicted that these small subsolar-mass neutron stars rapidly migrate inward due to a combination of gravitational-wave radiation and accretion torques, in an orbit-decay picture analogous to type-I and type-II migration in protoplanetary disks \citep{Ward:1997di, 2002ApJ...565.1257T, Lin:1986zz, Kley:2012ue}, leading to subsolar-mass mergers among multiple fragments \citep{Metzger:2024ujc, Lerner:2025dkd}, as well as mergers with the central black hole formed by stellar collapse \citep{Lerner:2025dkd}.
Observationally, the expectation is that these events give rise to multi-messenger events, including kilonova-like signatures inside supernovae \citep{Metzger:2024ujc, Kasliwal:2025keb}. AT2025ulz, discovered during follow-up of the August 18, 2025 sub-threshold trigger S250818k, sharpened interest in this scenario \citep{Kasliwal:2025keb, 2025arXiv251024620H,Franz:2025uan}, including the identification of a radio counterpart \citep{ODwyer:2026caq}, but see also subsequent Pan-STARRS, deep X-ray/radio, and HST follow-up for a stripped-envelope supernova interpretation \citep{2025ApJ...995L..27G, 2025ApJ...995L..47O, 2026ApJ...996L..24Y}.
In addition, there was a second recent subsolar-mass candidate event S251112cm for which EM follow-up searches have been carried out \citep{Vieira:2026eof}, and a potential association with a type IIb supernova has been proposed \citep{Hall:2026gov}.
In this work, we ask whether a significant gravitational-wave detection alone could determine the origin of a subsolar-mass neutron star. In particular, using scale-reduced numerical relativity simulations of hierarchical compact object mergers in a disk-like geometry, we demonstrate that multiple mergers can coherently impart kicks to the final binary, such that its orbit can attain substantial eccentricities in all cases considered. We argue that even for realistic separations these eccentricities will remain sizeable by the time of merger, making a potential eccentricity detection in a subsolar-mass gravitational-wave signal a clear signature of this formation channel. Establishing such a signature observationally is nevertheless challenging, because burst-like eccentric signals are difficult to search for, complete eccentric waveform models still cover only part of parameter space, and moderate eccentricity can be partly degenerate with spin precession in parameter inference \citep{2014PhRvD..90j3001T, 2017PhRvD..95b4038H, 2018PhRvD..98d4015H, 2023MNRAS.519.5352R}. In addition, we provide direct numerical-relativity waveforms for hierarchically merging systems and discuss their features.

\section{Basic picture}
\label{sec:picture}

In this work, we aim to model hierarchical mergers of subsolar-mass neutron stars in a collapsar disk scenario. In doing so, we will make a number of assumptions and simplifications that we list in the following.
Before doing so, we briefly summarize the current state in the literature of this model \citep{Piro:2006ja, Metzger:2024ujc, Chen:2025uwd, Lerner:2025dkd}.

In the collapsar disk scenario, we consider the death of a massive star with a rapidly rotating core, a comparatively rare outcome given the efficient angular-momentum transport expected in stellar interiors \citep{Fuller:2019ckz, Ma:2019cpr}, and the inferred rarity \citep{Cantiello:2007yy} of rapidly rotating engine-driven explosions \citep{Woosley:1993wj, Meszaros:2006rc, 2022ApJ...933L...9G}. The collapse of such a core (and failure to explode directly) will result in the formation of a rapidly spinning black-hole--disk system \citep{Woosley:1993wj, 1999ApJ...524..262M, Meszaros:2006rc}. This system is commonly invoked as the engine of most long-duration gamma-ray bursts \citep{Woosley:1993wj, 1999ApJ...524..262M, Meszaros:2006rc, 2022MNRAS.510.4962G, 2022ApJ...933L...9G, 2023ApJ...952L..32G}. The same engine has also been discussed as a potential source of r-process nucleosynthesis \citep{Siegel:2017nub, Agarwal:2025gbw} (though open questions remain \citep{Agarwal:2025gbw, Siegel:2017nub, Saleem:2025bme}). Black-hole masses can then be substantial, ranging from $M_{\rm BH} = 3-20 \, M_\odot$. The resulting accretion disk is very dense in the inner regions, $\rho >  10^{10}\,{\rm g\,cm^{-3}}$ \citep{Chen:2006rra, Piro:2006ja}, enabling rapid pair capture and neutronization \citep{Beloborodov:2002af, De:2020jdt}. In particular, if positron formation is suppressed \citep{Beloborodov:2002af, Chen:2006rra}, electron capture, $e^- +p \rightarrow n + \nu_e$, can dominate both the compositional evolution of the disk and its cooling, in addition to the dissociation of any $\alpha$ particles inherited from the progenitor star \citep{Chen:2006rra, Piro:2006ja}. Overall, this leads to a neutrino-dominated accretion flow (NDAF) with moderately thin scale height $h/r \simeq 0.1-1$ \citep{Chen:2006rra, Piro:2006ja}. It has then been argued that, given sufficiently high mass accretion rates, $\dot{M} \gg 0.5 \, M_\odot/s$, the outer $100 r_g$ of the disk can become gravitationally unstable to local collapse \citep{Piro:2006ja, Metzger:2024ujc}, with sufficient cooling provided by the processes above~\citep{Piro:2006ja, Chen:2006rra}, potentially leading to fragmentation of the disk with Jeans masses $M_J\simeq 0.02 \left(M_{\rm BH}/3M_\odot\right) (3h/r)^2$~\citep{Piro:2006ja, Metzger:2024ujc}.
Different works, both numerical \citep{Chen:2025uwd, Metzger:2024ujc} and analytical \citep{Lerner:2025dkd, Piro:2006ja}, have argued that this will lead to the rapid growth of $m=0.1\, M_\odot$ fragments, either by direct collapse or by rapid mergers of smaller fragments, at the expense of rapidly clearing the disk \citep{Lerner:2025dkd, Chen:2025uwd}. The resulting essentially gravitationally driven inspiral of the final fragment(s) is expected to take at most $\tau_{\rm GW} \simeq 7 \times 10^3\, {\rm s}\, (r/200\,r_g)^4(M_{\rm BH}/3\,M_\odot)^3 (m/0.1\,M_\odot)^{-1}$ for a circular orbit at the fragmentation radius~\citep{Piro:2006ja, Peters:1964zz}, although for formation closer in and more massive fragments this can be as short as $10-100\,\rm s$ \citep{Metzger:2024ujc, Chen:2025uwd}. In the case that smaller clumps are formed first, $m=0.001-0.01\, M_\odot$ (it is an open question how long these sub-Chandrasekhar-mass objects can remain stable \citep{1989ApJ...339..318C,1991ApJ...369..422C}), the disk would not be consumed by fragmentation itself, and the initial fragments would need to grow through an initial Bondi--Hoyle--Lyttleton accretion phase \citep{Bondi:1944rnk, Bondi:1952ni, Edgar:2004mk, Kaaz:2022dsa, Kim:2024zjb}, followed by type-I migration \citep{Ward:1997di, 2002ApJ...565.1257T, Goldreich:1980wa, Kley:2012ue} inside the disk \citep{Lerner:2025dkd}, and possibly, for $M_{\rm BH} \gtrsim 13 M_\odot$, gap opening and type-II migration within the inner $40\,r_g$ \citep{Lin:1986zz, Kley:2012ue, Lerner:2025dkd, Piro:2006ja}. While the two pictures differ in detail, both predict hierarchical multi-messenger mergers of fragments \citep{Metzger:2024ujc, Lerner:2025dkd}, with potentially kilonova and gamma-ray burst signatures from the merger \citep{Metzger:2024ujc, Kasliwal:2025keb}, or even flare activity or prolonged jet activity from the migration itself \citep{Lerner:2025dkd}.
While each of these scenarios merits proper and full investigation, we pursue a different but simpler question. Since associating and interpreting these electromagnetic transients may be ambiguous, we instead ask whether there are additional signatures, not tied to electromagnetic emission, that could identify this scenario.
Motivated by the fact that substantial eccentricities can be excited both in hierarchical black-hole mergers in clusters \citep{Samsing:2017xmd} and in giant-impact mergers during planet formation \citep{2001Icar..152..205C, 1993Icar..106..247G, 2015ApJ...807..157T}, we investigate the late inspiral and merger of several fragments using full numerical relativity. In particular, we focus on fragments that have inspiralled or migrated to the inner $20\, r_g$, consistent with the picture outlined in \citet{Metzger:2024ujc} and \citet{Lerner:2025dkd}. Ideally, larger scale separations should also be considered, but doing so is computationally challenging in numerical relativity. Finally, since we are primarily interested in modeling orbital trajectories and gravitational waveforms, which to lowest order do not depend strongly on the matter content of the spacetime, we model all fragments as black holes for computational efficiency. We caution, however, that collisions involving neutron stars, especially outside of binary orbits, are accompanied by substantial mass loss \citep{East:2011xa, East:2012ww, Chaurasia:2018zhg, Papenfort:2018bjk}, which could alter some of the detailed scattering and merger dynamics considered here. Matter-coupled subsolar-mass merger simulations of circular binaries have also recently begun to be explored \citep{Markin:2023fxx, Corman:2026lbt}.

\begin{figure*}
    \centering
    \includegraphics[width=1.0\linewidth]{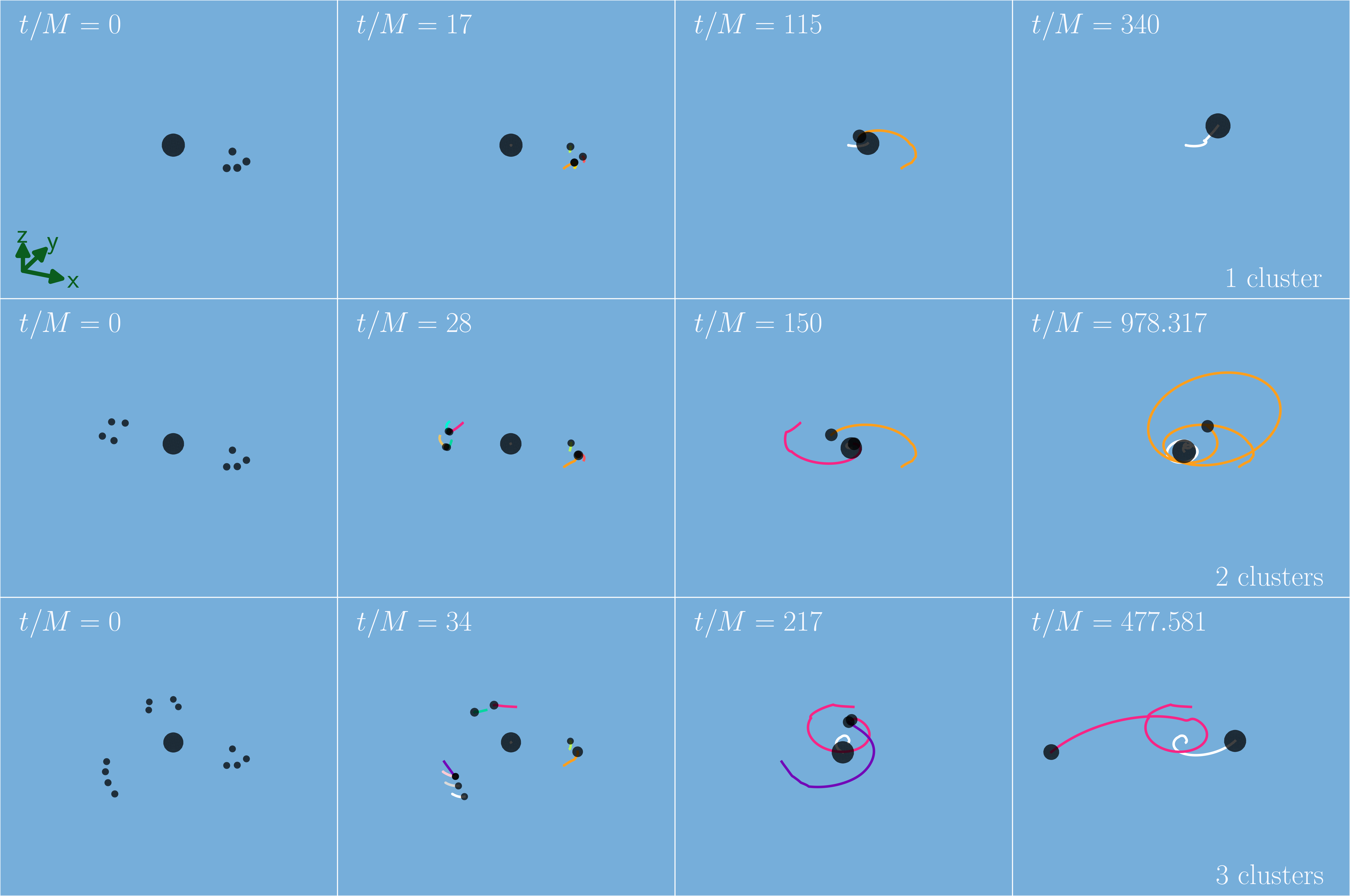}
    \caption{Fragment merger dynamics in a collapsar-disk-like arrangement. Shown are three different configurations, consisting of 1 initial cluster (top row), 2 clusters (middle row), and 3 clusters (bottom row). Each panel shows the orbital tracks accumulated up to the time \(t/M\) = $ct/r_g$ printed in the upper-left corner, using the same viewing angle in all panels. 
    Here $r_g = GM/c^2$, where $M$ is the total initial mass of the system.
    The dynamics occur almost entirely in the midplane. Colored curves trace the previous motion of the respective compact objects, while black spheres mark their instantaneous positions; the sphere radii are scaled by mass 
    to visualize the changing masses after mergers.}

    \label{fig:3D_render}
\end{figure*}

\section{Methods}

The hierarchical subsolar-mass compact-object mergers are set up as a sequence of local clusters, each containing four black holes. While the initial bulk motion of each cluster is assumed to be Keplerian (the details of which in reality will depend on the presence of an accretion disk and the relevance of migration torques \citep{Lerner:2025dkd, Piro:2006ja}), we include an effective velocity dispersion 
\begin{equation}
    P(u_R, u_\phi, u_z)=\frac{1}{(2\pi)^{3/2}\sigma}\text{exp}\left[-\frac{u_R^2}{2\sigma_R^2}-\frac{u_\phi^2}{2\sigma_\phi^2}-\frac{u_z^2}{2\sigma_z^2}\right],
\end{equation}
where $\boldsymbol{u}=\boldsymbol{v}-v_\text{kep}\hat{\boldsymbol{\phi}}$ is the random-motion component of the velocity, $v_\text{kep}$ the local Keplerian velocity, and $\boldsymbol{v}$ the velocity of each compact object,  $\sigma = \sigma_R\sigma_\phi \sigma_z$, and $\sigma_R,\sigma_\phi,\sigma_z$ are the dispersion widths in the three directions. Here we adopt the three coplanar initial configurations summarized in Fig.~\ref{fig:3D_render} and listed explicitly in Appendix~\ref{app:initial_data} (Table~\ref{tab:initial_punctures}), consisting of one, two, or three four-body clusters orbiting a central black hole.

For reasons of computational feasibility, we are unable to model large mass ratios, $q$, relative to the central black hole mass. As a result, we fix the central black hole to have mass $M_{\rm BH} = 3\,M_\odot$, and adopt a rapid dimensionless spin $\chi\simeq0.52$.
Each cluster then consists of four equal-mass black holes in bare puncture mass, adding up to a total mass of $M_{\rm cluster} \simeq 0.8\,M_\odot$, which fixes the cluster-to-total mass ratio $q \equiv M_{\rm cluster}/(M_{\rm cluster} + M_{\rm BH}) \approx 0.21$.
The subsolar-mass compact objects adopted here have aligned
dimensionless spins $\chi\sim0.5$, in part motivated by prior mergers,
though non-spinning fragments should also be considered in future work.

Numerically, we model the black holes as punctures \citep{Brandt:1997tf}, generated using the \textit{SolvePunctures} module of the \texttt{SpECTRE} numerical relativity code \citep{Kidder:2016hev, Vu:2021coj, Vu:2024cgf, spectrecode}. The initial puncture masses, coordinates, and coordinate velocities used for all three runs are collected in Appendix~\ref{app:initial_data} (Table~\ref{tab:initial_punctures}). The initial data are then interpolated from the spectral grid onto a finite-difference domain in the performance-portable \texttt{AthenaK} code \citep{2026ApJS..283...27S}. We then evolve the spacetime using the Z4c formulation of dynamical spacetime \citep{Hilditch:2012fp} within the \texttt{AthenaK} dynamical-spacetime solver \citep{Zhu:2024utz}. We adopt moving-puncture gauges \citep{Alcubierre:2002kk, Campanelli:2005dd}, using a spatially varying Gamma-driver damping following \citet{Bamber:2025gxj}, as well as spatially varying constraint damping appropriate to the large mass hierarchy in the system.


\begin{figure*}
    \centering
    \begin{minipage}[t]{0.42\textwidth}
        \centering
        \includegraphics[width=\linewidth]{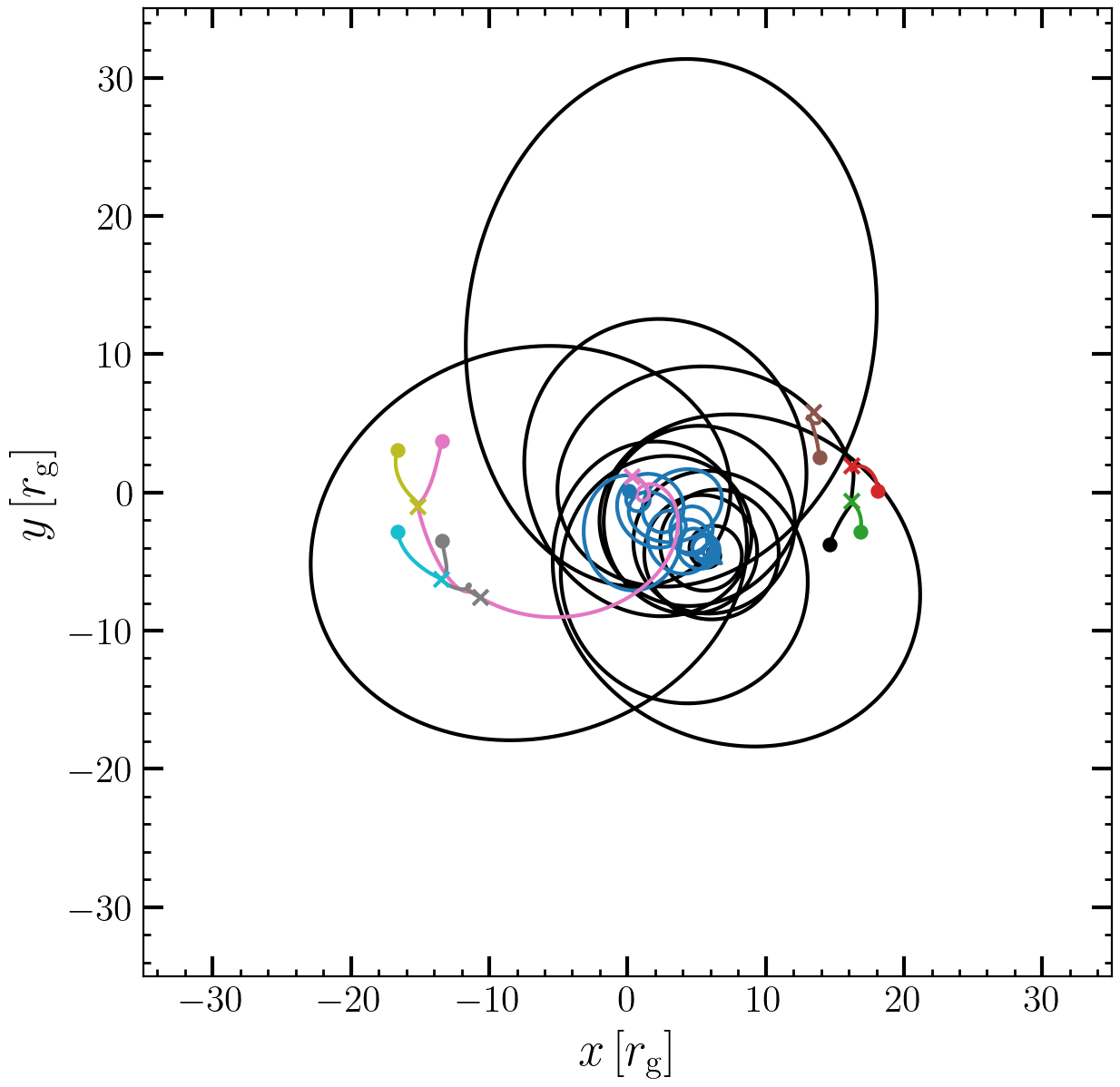}
    \end{minipage}\hfill
    \begin{minipage}[t]{0.57\textwidth}
        \centering
        \includegraphics[width=\linewidth]{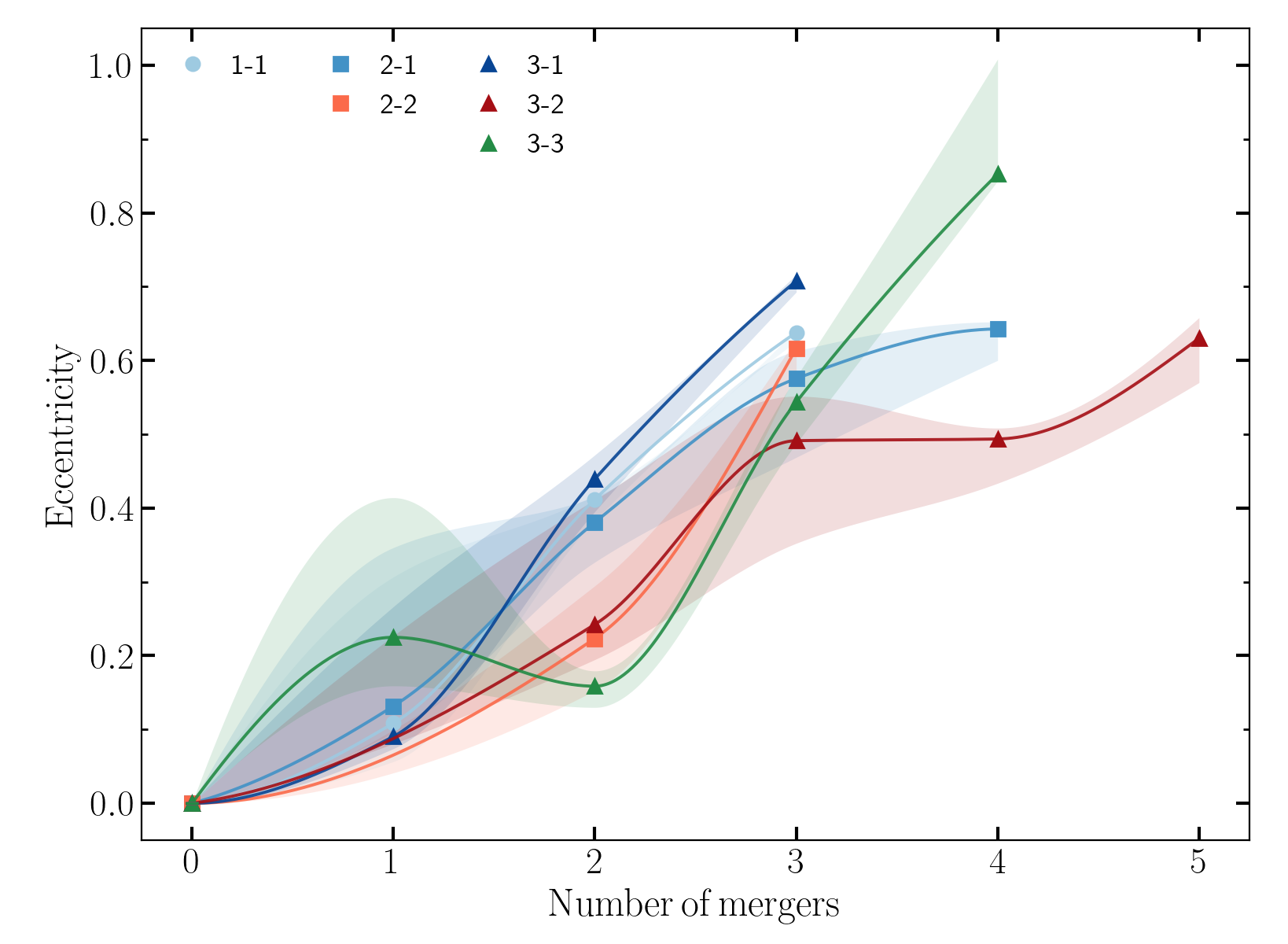}
    \end{minipage}
    \caption{{\it (Left)} compact-object trajectories in the orbital \(xy\)-plane for the two-cluster simulation. Each colored track follows a compact object from its initial position, marked by a circle, to its last recorded position. Crosses mark tracks that terminate in mergers, while upward triangles mark the two remnants that survive to the final simulation time. Trajectories are shown in the inertial coordinate frame. {\it (Right)} representative eccentricity histories for the merger lineages in the three runs, plotted against merger stage. The six series labeled 1-1, 2-1, 2-2, 3-1, 3-2, and 3-3 correspond to the tracked remnants in the one-, two-, and three-cluster configurations. Symbols mark the median orbital eccentricity measured after each merger event, the solid curves are smooth interpolants through those medians, and the shaded bands show the 1.5-$\sigma$ uncertainty percentile range within each window.}
    \label{fig:traj_ecc}
\end{figure*}

\section{Results}

We now discuss the results of our hierarchical merger simulations in a collapsar-disk-like scenario.
Our simulations begin after multiple fragments have migrated inward, depending on their initial masses, either through gravitational-wave radiation or under the combined action of migration torques (see Section~\ref{sec:picture}). We consider three different cluster configurations. Starting from the initial state, the evolution at representative times is shown in Fig.~\ref{fig:3D_render}. The evolution consists of three stages: initial fragment merger, formation of a large subsolar-mass neutron star, and its eventual merger with the central black hole.
Initially, each local cluster contains a set of individual subsolar-mass compact objects (left panel, Fig.~\ref{fig:3D_render}). Shortly after the start of the simulation, the smaller fragments begin to collide. None of the collisions we observe lead to circularization of an intermediate binary in any of the simulations. Instead, we find that the black holes get gravitationally focused and merge with the innermost black hole of each cluster. 
We further analyze this behavior in the case of the two-cluster system by comparing the different orbital trajectories (Fig.~\ref{fig:traj_ecc}, left panel).
Here we can clearly identify the three initial stages, as well as the large kicks imparted to the orbit of the final binaries.
This initial phenomenon can be understood in terms of Newtonian planetary dynamics, which we present for the same initial conditions in Appendix~\ref{app:Newtonian}. In the absence of gravitational-radiation reaction, circular orbits for planetary motion will be stable if the planets stay outside of the mutual Hill radius $r_H$ \citep{1993Icar..106..247G}. If this is not the case, orbits will develop a net eccentricity and begin to cross. Based on our initial conditions, the initial orbit has both a net eccentricity $e_{\rm initial} \simeq 0.1$ and overlapping mutual Hill spheres.
In particular, the minimum initial eccentricity for fragment merger (for two orbits of equal eccentricity, as in our case) can be approximated as $e_{\min} \simeq 0.5 \Delta a/a$, where $\Delta a$ is the separation between the two subsolar-mass black holes \citep{2015ApJ...807..157T}. We therefore find that, characteristically for our initial setup, $e_{\rm initial} \simeq 0.1 > 0.05 \simeq e_{\min}$.
Accordingly, a comparison Newtonian N-body simulation (see Appendix Fig.~\ref{fig:Newtonian}) shows similar merger trajectories and dynamics inside each cluster. This behavior bears some resemblance to the late-stage giant-impact phase of terrestrial-planet formation~\citep{2001Icar..152..205C}.

Each merger imparts a net kick $v_{\rm kick}$ onto the growing fragment as
discussed above. We observe this directly as kinks in the orbital
trajectories shown in Fig.~\ref{fig:3D_render}. While the initial mergers
within each cluster are rapid, subsequent mass growth leads to the
formation of a final binary or triple system (in the case of three initial
clusters).  This behavior is very different from the Newtonian case, as
radiation-reaction-driven shrinkage of the orbit and Post-Newtonian
dynamics dominates. Indeed, by comparing to the Newtonian evolution
(Appendix Fig.~\ref{fig:Newtonian}), we can see that the
general-relativistic case features rapid mergers, whereas the Newtonian
case features multiple scattering events, the build-up of highly eccentric
orbits with large apsidal precession, and the ejection of unbound fragments
at the end of the integration.

While the overall dynamics is the same in the case of three initial
clusters, we point out that the repeated merger of all clusters, leads to
the formation of a final solar-mass secondary. As such, the collapsar scenario can, in principle, also produce eccentric solar-mass neutron star -- black hole events, such as GW200105 \citep{Morras:2025xfu}, although a triple formation channel appears more likely \citep{Stegmann:2025clo}.

\begin{figure*}
    \centering
    \includegraphics[width=1.0\linewidth]{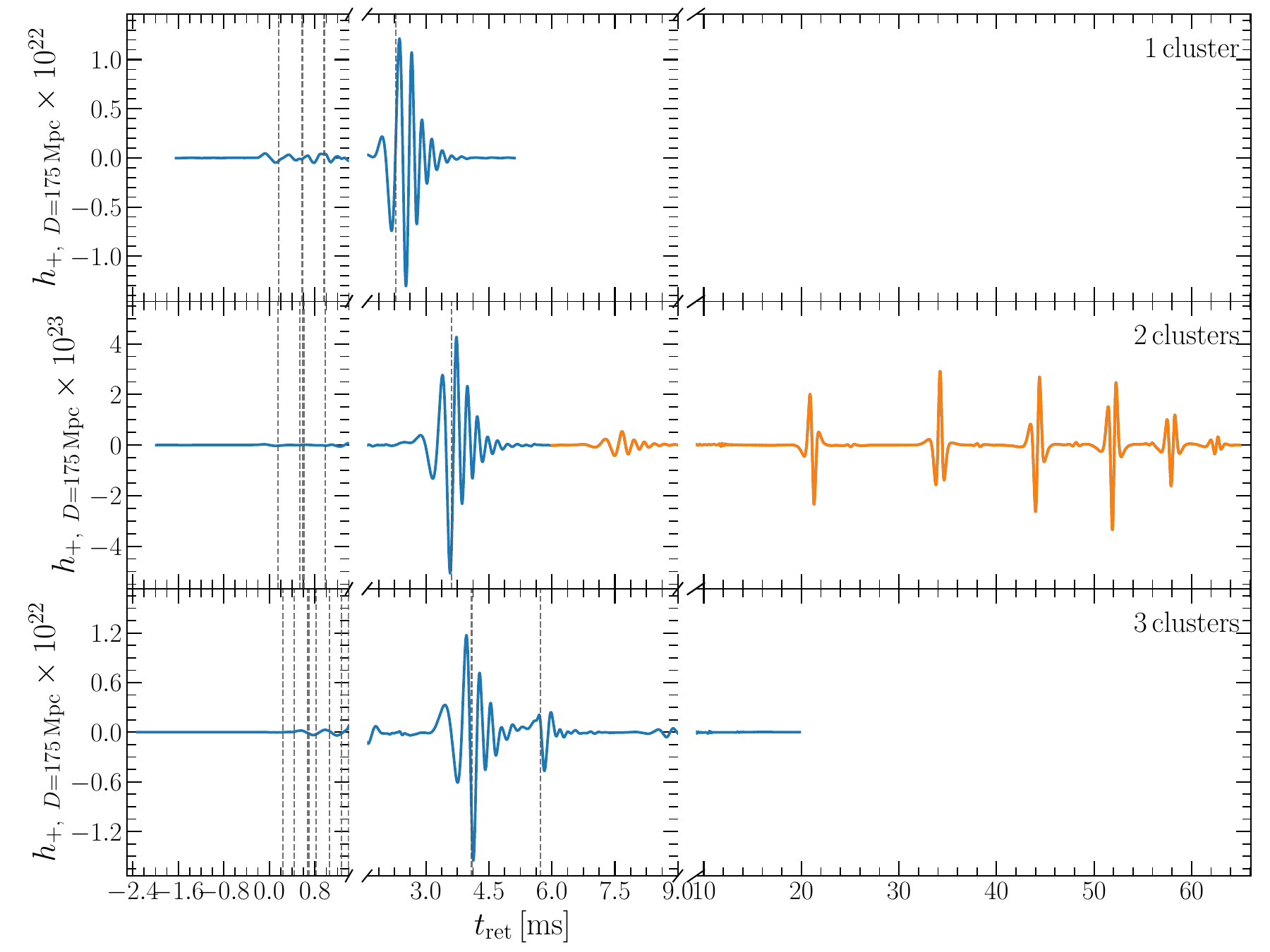}
    \caption{Detector-frame plus-polarization strain $h_+$ for the three simulations, reconstructed from the extracted waveform at $r=80\,M$ and scaled to a fiducial distance of $D=175\,\mathrm{Mpc}$. The horizontal axis shows retarded time and uses a broken layout to separate the early, intermediate, and late merger episodes. 
    Vertical dashed lines mark the refined merger times during the
    evolution, orange color denotes the final eccentric inspiral. Initially the merger of smaller fragments proceeds rapidly, followed either by the final merger in the case of 1 cluster, or by the
formation of the final binary system, consisting of the primary central
black hole and the hierarchically assembled secondary, which is
subsolar-mass in the case of the two cluster simulation. This inspiral is
highly eccentric, as can be seen from the pulse-like waveform signa0l prior
to the final merger. The three cluster simulation has a final orbital
period longer than what we simulate here, and the single cluster system has
completely merged already.}
    \label{fig:waveform}
\end{figure*}

After this hierarchical merger stage, the interaction with the massive
primary (central black hole), with which the subsolar-mass companion will
ultimately merge, becomes important. Because we do not include misaligned
spins, the dynamics remain in the orbital plane. Finally, this binary will
begin to circularize and merge rapidly, while still exhibiting visibly
non-circular (eccentric) orbits by the time of the final inspiral. 

After each fragment merger, we observe an increase in individual orbital
eccentricity. We then proceed to quantify this as follows. We extract the
trajectories of the surviving black hole in each cluster (i.e., the one
that undergoes growth through repeated capture of and merger with smaller
fragments). While eccentricity is generally a non-trivial quantity to
define in numerical relativity simulations (see  \citet{Shaikh:2023ypz,
Boschini:2024scu} for different definitions), we are here primarily
concerned with quantifying the build-up of potentially large eccentricity
$e>0.1$. To demonstrate this, we adopt a simple Newtonian
coordinate-eccentricity proxy computed from the compact-object tracker
positions and velocities relative to the current most massive active
puncture.  In particular, for a fragment or cluster center of mass with
mass $m$ orbiting a central puncture of current mass $M_{\rm c}$, we use
the instantaneous eccentricity vector (Laplace--Runge--Lenz vector)
$\vec{e} = \vec{v}_{\rm rel} \times (\vec{r}_{\rm rel} \times \vec{v}_{\rm
rel})/[G(M_{\rm c}+m)] - \hat{r}_{\rm rel}$, where $\vec{r}_{\rm rel}$ and
$\vec{v}_{\rm rel}$ are measured from the central puncture to the fragment
or cluster center of mass. We can then recover an estimate for the orbital
eccentricity $e=|\vec{e}|$; see Fig.~\ref{fig:traj_ecc} (right panel),
where we can see a clear trend of increasing eccentricity during the first
mergers in each cluster. Since eccentricity growth is dominant during the
first few mergers that have Newtonian analogues, we can estimate the
eccentricity increment per merger from giant-impact-style gravitational
focusing \citep{Lissauer:1993dr, 2001Icar..152..205C, 2015ApJ...807..157T}.
For our setup, $\Delta a / a \sim 0.1$, and the eccentricity kick imparted
by the $k$th merger we observe approximately scales as $\Delta e_k \sim
\Delta a/a \sim 0.02-0.1$, not unlike values observed in studies of
giant-impact scenarios \citep{2006ApJ...642.1131K}, where the exact
analytic prefactor depends on the dynamics of the collisions (e.g.,
Appendix of \citet{2001Icar..150..303F}).

\begin{figure}
    \centering
    \includegraphics[width=1.0\linewidth]{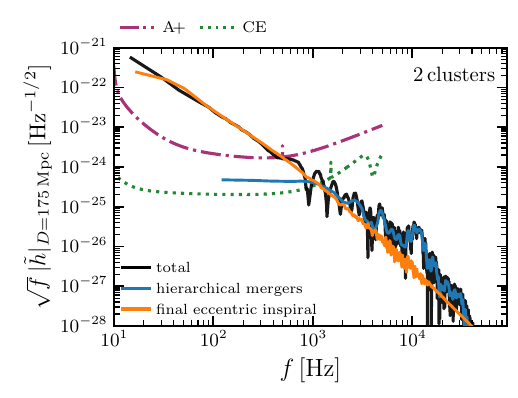}
    \caption{Frequency-domain detector-frame gravitational-wave spectrum
      $\sqrt{f}\,|\tilde h|$ at $D=175\,\mathrm{Mpc}$ for the two-cluster
      simulation extracted at $r=80\,r_g$. The black curve shows the
      full-signal spectrum, and the colored solid curves correspond to the
      hierarchical fragment merger and final merger with the central black
      hole, respectively. The split between these two windows 
      is shown in Fig.~\ref{fig:waveform}. The magenta and green curves show the LIGO A+ and Cosmic Explorer sensitivity amplitudes.}
    \label{fig:spectrum}
\end{figure}

In fact, we would expect the merger of smaller fragments generically to induce eccentricity in this system. In reality, probing these large eccentricities for a $3\,M_\odot$ black hole leads to a number of interesting consequences.
First, while gravitational-radiation reaction ultimately circularizes the orbit, along Peters trajectories the semimajor axis and eccentricity are linked by $a \propto e^{12/19}$ \citep{Peters:1964zz}, so that to leading order $\dot{e}/e \approx (19/12)\, \dot{a}/a$. The orbit therefore circularizes only modestly faster than it shrinks (see Fig.~\ref{fig:traj_ecc}).
A substantial residual eccentricity therefore persists into the detector band, as quantified in Section~\ref{sec:gw-emission}. In the neutron-star case, at sufficiently close separations, pericenter approaches the tidal-disruption radius and the fragment may begin Roche-lobe filling during late pericenter passages, leading to mass transfer that will likely regulate the fragment mass to remain below $1\,M_\odot$ for low-mass central black holes \citep{Zenati:2024elj}.
This episodic fueling of the black hole may also increase the chances of producing flare-like activity in these systems \citep{Perna:2005tv, Piro:2006ja}.

\subsection{Gravitational-wave emission}
\label{sec:gw-emission}
To better understand the dynamics of gravitational-wave emission in this regime, we now analyze both the waveforms and the gravitational-wave spectra. 
We focus specifically on the two-cluster run as a fiducial configuration. 
To do so, we extract the gravitational-wave strain using fixed-frequency integration \citep{Reisswig:2010di} at a distance of $r=80\, r_g$ from the center. The resulting time-domain waveforms are shown in Fig.~\ref{fig:waveform}. We can identify several phases. First, we can see that multiple mergers of subsolar-mass black holes dominate the early signal within the first millisecond of the simulation. While this stage is arguably strongly dependent on our initial conditions, it does set the eccentricity conditions for subsequent mergers as discussed above. As expected from the mass scaling of the frequency as well as the effective capture dynamics of the mergers, the characteristic frequencies of this phase remain around $10\,{\rm kHz}$ (see the blue line in Fig.~\ref{fig:spectrum}). In all systems, this is followed either by the main merger (one cluster) or by a stronger burst due to the merger of the two or three remaining fragments (two- and three-cluster configurations, respectively). We find that this signal is burst-like, as almost no inspiral component is present, and has support above $1\,{\rm kHz}$.
As we can see, all of this hierarchical merger dynamics is too high in frequency and too low in amplitude for a fiducial distance of $175$ Mpc to be detectable with LIGO at A+ sensitivity \citep{KAGRA:2013rdx,LIGOScientific:2016wof}, and is barely detectable (and only partially) with Cosmic Explorer \citep{Evans:2021gyd}.
We caution, however, that for subsolar mass mergers on larger orbits and separations, there could longer inspiral dynamics leading to potentially detectable mergers of two sub-solar mass fragments, e.g., as in candidate event S251112cm.
In either case, the final inspiral of the hierarchically merged massive, yet still subsolar-mass\footnote{In the two cluster case, the initial clusters (with mass $0.8\, M_\odot$ initially) merge with the central black hole individually, making each merger subsolar-mass in nature.} fragment undergoes several inspiral orbits at high eccentricity $(e \simeq 0.6)$, e.g., for the two-cluster scenario (Fig.~\ref{fig:waveform}). Indeed, given the masses present and the inspiral frequencies around $100\,$Hz, we find that this signal has substantial support in the LIGO A+ band (black curve in Fig.~\ref{fig:spectrum}). The parameters of such a signal can be estimated by considering the mass and spin of the system at this point. Specifically, we adopted an isolated horizon formalism \citep{Dreyer:2002mx} to extract the individual masses and spins of the final merging system, which in reality would be a black hole -- neutron star system. Using the isolated horizon diagnostics as implemented in \texttt{SpECTRE} (based on \citet{Gundlach:1997us}), which we have also compared to those measured with the implementation in the \texttt{BHaHAHA} code ( \citet{etienne2026bhahahafastrobustapparent}), we obtain a mass pair of $(m_1; \chi_{1})-(m_2; \chi_{2})$ corresponding to $(3.71 M_\odot;0.76)-(0.80M_\odot; 0.66)$ for the two cluster, and $(3.73M_\odot;0.71)-(1.57M_\odot; 0.34)$ for the three cluster system. We caution that these values should be taken with a grain of salt. While the primary spins are close to the Keplerian value for a rapidly rotating neutron star (e.g., \citealt{Cook:1993qr,Breu:2016ufb,Koliogiannis:2019rvh}), the primary black hole at this point might have efficiently spun down by the strong jet launched from the accretion flow \citep{Jacquemin-Ide:2023aax}.

To assess how much of this eccentricity survives down to merger, we integrate the Peters equations~\citep{Peters:1964zz} in $(a, e)$ from a range of initial separations $a_0$ down to the innermost stable circular orbit. For our two-cluster fiducial configuration with $e_0 \simeq 0.6$, starting from $a_0 \sim 50\,r_g$ leaves $e \simeq 0.47$ at $f_{\rm GW} = 100\,{\rm Hz}$ and $e \simeq 0.12$ at $500\,{\rm Hz}$, well within the detectable regime \citep{Morras:2025xfu}. Starting from the disk fragmentation radius $a_0 \sim 100\,r_g$ \citep{Piro:2006ja, Metzger:2024ujc,Lerner:2025dkd}, Peters circularization reduces these values to $e \simeq 0.2$ at $100\,{\rm Hz}$ and $e \simeq 0.05$ at $500\,{\rm Hz}$. The hierarchical-merger signature is therefore most pronounced in the early in-band inspiral ($\sim$ tens to hundreds of Hz) rather than near merger, and is clearest for loud nearby events or at next-generation detector sensitivity~\citep{LIGOScientific:2016wof}. This is fully consistent with the expectation that only the final merger in this scenario will be observable by LIGO \citep{Chen:2025uwd}.
One interesting effect, not captured here because of the black-hole nature of our setup, is the excitation of f-modes at pericenter passage \citep{Chirenti:2016xys}. This has been described both analytically \citep{Lai:1993di, Kokkotas:1995xe} and numerically \citep{Gold:2011df, Chaurasia:2018zhg, Rosofsky:2018vyg, Papenfort:2018bjk}, and is expected to potentially power multi-messenger signals \citep{Tsang:2011ad, Tsang:2013mca, Neill:2021lat}. For the mass ranges considered here, the final neutron-star--black-hole merger may also undergo mass transfer, which could naturally regulate the mass of the fragment (below $1\, M_\odot$) and power flare-like transients at every pericenter passage \citep{Zenati:2024elj}.

\section{Conclusions}

We have studied hierarchical mergers of subsolar-mass fragments in a collapsar-disk-like scenario modeled by scale-reduced numerical relativity simulations of up to 13 merging black holes. Across the different initial conditions and scenarios considered here, repeated capture and merger events generically build up substantial orbital eccentricity in the final surviving fragment. In our simulations, this growth reaches values as large as $e \sim 0.6$ (though the precise details will depend on the merger history), consistent with the expectation from repeated kick accumulation and gravitational focusing during the hierarchical assembly phase \citep{2001Icar..150..303F, 2006ApJ...642.1131K}. We validate this initial picture by performing a direct comparison with Newtonian N-body dynamics calculations, supporting the interpretation that the first stages of this dynamics are closely related to giant-impact-style orbital crossing and mergers \citep{1993Icar..106..247G, 2001Icar..152..205C, 2015ApJ...807..157T}.

From the gravitational-wave point of view, our results suggest a clear
separation between the unobservable fragment merger dynamics and the
potentially observable final (subsolar-mass) merger with the central black
hole. The early hierarchical mergers occur at frequencies around 10 kHz and
low strain amplitudes, far outside the regime of current and planned
gravitational-wave detectors. By contrast, the final merger between the
central black hole and the hierarchically assembled secondary can enter the
ground-based detector band with appreciable residual eccentricity not
unlike eccentricity values claimed for the neutron-star--black-hole merger
GW200105 \citep{Morras:2025xfu}, although a merger in a hierarchical triple system seems most likely for that event \citep{Stegmann:2025clo,Romero-Shaw:2025otx,Morras:2026mrv}. Within our most massive simulation,
hierarchical mergers lead to the formation of a solar-mass secondary.
In general, for a black hole in a collapsar primary, we point out that consistent with the expectation that rapid
mass accretion necessary for subsolar mass formation \citep{Lerner:2025dkd} in a magnetically arrested regime \citep{Narayan:2003by,Tchekhovskoy:2011zx} leads to the
rapid spin down of the central black hole \citep{Lowell:2023kyu,Lowell:2025ris},
which in the collapsar scenario can be around $a \simeq 0.2$ \citep{Jacquemin-Ide:2023aax},
consistent with the expectation of low spin in black hole neutron star mergers, incl. GW200105 \citep{Morras:2025xfu}.
Especially, if such a system were observed as a subsolar-mass
compact-binary candidate, a robust measurement of eccentricity would
therefore be a strong indication for a hierarchical formation channel in a
collapsar disk rather than an isolated binary origin.

At the same time, our present calculations should be viewed as a proof of principle. We have modeled the fragments as black holes, adopted reduced scales and moderate mass ratios for computational tractability, and not included the hydrodynamic response of neutron-star matter \citep{East:2011xa,Papenfort:2018bjk} or the full disk environment. In particular, the absence of gas in our simulations means that disk-migration torques~\citep{Piro:2006ja, Lerner:2025dkd} and dynamical friction from a population of smaller, unmerged fragments~\citep{Metzger:2024ujc} are not modeled. Gas may also damp or excite eccentricity through disk torques \citep{Papaloizou:2000, Goldreich:2003} and modify spins through accretion of ambient gas with finite angular momentum \citep{Li:2022ApJL928L1, Chen:2022ApJL939L23}. It will also be important to properly understand gravitational instability and initial fragment formation \citep{Chen:2025uwd} in full general relativity, as well as early merger channels at separations of $50-100\,r_g$ to understand the resulting merger dynamics and its impact on the final merger.

Finally, performing full numerical relativity simulations coupled to matter evolution \citep{Markin:2023fxx,
Markin:2026eyc, Corman:2026lbt} will be important to clarify potential flare and merger-like transients from the final eccentric neutron-star--black-hole merger \citep{Zenati:2024elj}, which would lend additional predictability to this scenario.

\section{Acknowledgments}
The authors are grateful for discussions with Yixian Chen, Yuri Levin, Daniel Kasen, Anthony Readhead, Haiyang Wang, Alan Weinstein, and Yossef Zenati, and to Hengrui Zhu for help with initial data import.
ERM and JW acknowledge support from NASA's ATP program under grant 80NSSC24K1229 and from the National Science Foundation under grant No. PHY-2309210.
NV and KN acknowledge support by the National Science Foundation through grants No.~PHY-2309211; No.~PHY-2309231;  No.~OAC-2513339 at Caltech; and NASA award No.~80NSSC26K0340, and ND, LK and WT through awards No.~PHY-2407742; No.~PHY-2207342; No.~OAC-2513338; and NASA award No.~80NSSC26K0340 at Cornell. NV, ND, LK, KN, and WT acknowledge support by the Sherman Fairchild Foundation at Caltech and Cornell.
    Simulations were carried out on AMD's AI \& HPC cluster.
	Additional simulations were also performed on the NSF Frontera supercomputer under grant AST21006.
    This work also made use of Delta at the National Center for Supercomputing Applications (NCSA) through allocation PHY210074 from the Advanced Cyberinfrastructure Coordination Ecosystem: Services \& Support (ACCESS) program, which is supported by National Science Foundation grants \#2138259, \#2138286, \#2138307, \#2137603, and \#2138296.  This research was supported in part by grant NSF PHY-2309135 to the Kavli Institute for Theoretical Physics.
\appendix

\section{Initial Conditions}
\label{app:initial_data}
The initial parameters for the punctures are listed in Table~\ref{tab:initial_punctures}.

\begin{table*}[t]
    \centering
    \scriptsize
    \caption{Initial puncture coordinates $(x,y,z)$ and coordinate velocities $(v_x, v_y, v_z)$ for all three simulated configurations. The mass column lists the initial apparent-horizon Christodoulou masses with the central black hole in each run having physical mass $3\,M_\odot$. The dimensionless spins $\chi$ are the corresponding apparent-horizon spin magnitudes. The \texttt{MBH\_3\_1} and \texttt{MBH\_3\_2} runs correspond to the puncture subsets shown here for the full \texttt{MBH\_3\_3} realization.}
    \label{tab:initial_punctures}
    \resizebox{\textwidth}{!}{%
\begin{tabular}{c c c c c c c c c c c}
        \hline
        Config. & ID & Cluster & $M_{\rm Ch}\,[M_\odot]$ & $x$ & $y$ & $z$ & $v_x$ & $v_y$ & $v_z$ & $\chi$ \\
        \hline
        \multicolumn{11}{c}{\texttt{MBH\_3\_1}} \\
        \hline
        MBH\_3\_1 & 0 & -- & 3.000000 & 0.100000 & 0.100000 & 0 & 0 & 0 & 0 & 0.520 \\
        MBH\_3\_1 & 1 & 1 & 0.199645 & 14.588887 & -3.782286 & 0 & $1.57\times 10^{-1}$ & $4.24\times 10^{-1}$ & $-6.45\times 10^{-3}$ & 0.446 \\
        MBH\_3\_1 & 2 & 1 & 0.198897 & 16.841732 & -2.852019 & 0 & $8.99\times 10^{-2}$ & $3.99\times 10^{-1}$ & $-5.03\times 10^{-3}$ & 0.439 \\
        MBH\_3\_1 & 3 & 1 & 0.201558 & 18.100000 & 0.100000 & 0 & $-2.39\times 10^{-2}$ & $4.19\times 10^{-1}$ & $1.19\times 10^{-2}$ & 0.467 \\
        MBH\_3\_1 & 4 & 1 & 0.207106 & 13.887309 & 2.531074 & 0 & $-5.64\times 10^{-2}$ & $4.70\times 10^{-1}$ & $-2.64\times 10^{-3}$ & 0.518 \\
        \hline
        \multicolumn{11}{c}{\texttt{MBH\_3\_2}} \\
        \hline
        MBH\_3\_2 & 0 & -- & 3.000000 & 0.100000 & 0.100000 & 0 & 0 & 0 & 0 & 0.517 \\
        MBH\_3\_2 & 1 & 1 & 0.198812 & 14.588887 & -3.782286 & 0 & $1.57\times 10^{-1}$ & $4.24\times 10^{-1}$ & $-6.45\times 10^{-3}$ & 0.453 \\
        MBH\_3\_2 & 2 & 1 & 0.200605 & 16.841732 & -2.852019 & 0 & $8.99\times 10^{-2}$ & $3.99\times 10^{-1}$ & $-5.03\times 10^{-3}$ & 0.453 \\
        MBH\_3\_2 & 3 & 1 & 0.199724 & 18.100000 & 0.100000 & 0 & $-2.39\times 10^{-2}$ & $4.19\times 10^{-1}$ & $1.19\times 10^{-2}$ & 0.472 \\
        MBH\_3\_2 & 4 & 1 & 0.206426 & 13.887309 & 2.531074 & 0 & $-5.64\times 10^{-2}$ & $4.70\times 10^{-1}$ & $-2.64\times 10^{-3}$ & 0.525 \\
        MBH\_3\_2 & 5 & 2 & 0.200039 & -13.422962 & 3.723467 & 0 & $-1.01\times 10^{-1}$ & $-4.67\times 10^{-1}$ & $2.20\times 10^{-2}$ & 0.460 \\
        MBH\_3\_2 & 6 & 2 & 0.200062 & -13.422962 & -3.523467 & 0 & $9.05\times 10^{-2}$ & $-4.52\times 10^{-1}$ & $4.24\times 10^{-3}$ & 0.470 \\
        MBH\_3\_2 & 7 & 2 & 0.208143 & -16.641732 & 3.052019 & 0 & $-1.27\times 10^{-1}$ & $-3.92\times 10^{-1}$ & $-6.00\times 10^{-5}$ & 0.531 \\
        MBH\_3\_2 & 8 & 2 & 0.200212 & -16.641732 & -2.852019 & 0 & $7.90\times 10^{-2}$ & $-4.38\times 10^{-1}$ & $-2.62\times 10^{-2}$ & 0.463 \\
        \hline
        \multicolumn{11}{c}{\texttt{MBH\_3\_3}} \\
        \hline
        MBH\_3\_3 & 0 & -- & 3.000000 & 0.100000 & 0.100000 & 0 & 0 & 0 & 0 & 0.515 \\
        MBH\_3\_3 & 1 & 1 & 0.197493 & 14.588887 & -3.782286 & 0 & $1.57\times 10^{-1}$ & $4.24\times 10^{-1}$ & $-6.45\times 10^{-3}$ & 0.453 \\
        MBH\_3\_3 & 2 & 1 & 0.199716 & 16.841732 & -2.852019 & 0 & $8.99\times 10^{-2}$ & $3.99\times 10^{-1}$ & $-5.03\times 10^{-3}$ & 0.432 \\
        MBH\_3\_3 & 3 & 1 & 0.198341 & 18.100000 & 0.100000 & 0 & $-2.39\times 10^{-2}$ & $4.19\times 10^{-1}$ & $1.19\times 10^{-2}$ & 0.471 \\
        MBH\_3\_3 & 4 & 1 & 0.205009 & 13.887309 & 2.531074 & 0 & $-5.64\times 10^{-2}$ & $4.70\times 10^{-1}$ & $-2.64\times 10^{-3}$ & 0.522 \\
        MBH\_3\_3 & 5 & 2 & 0.197178 & -3.523467 & 13.622962 & 0 & $-4.55\times 10^{-1}$ & $-1.46\times 10^{-1}$ & $2.20\times 10^{-2}$ & 0.453 \\
        MBH\_3\_3 & 6 & 2 & 0.198605 & -9.799495 & 9.999495 & 0 & $-3.46\times 10^{-1}$ & $-3.04\times 10^{-1}$ & $4.24\times 10^{-3}$ & 0.467 \\
        MBH\_3\_3 & 7 & 2 & 0.204268 & -5.714342 & 16.074775 & 0 & $-4.03\times 10^{-1}$ & $-8.60\times 10^{-2}$ & $-6.00\times 10^{-5}$ & 0.522 \\
        MBH\_3\_3 & 8 & 2 & 0.198571 & -10.827389 & 13.122756 & 0 & $-3.40\times 10^{-1}$ & $-2.87\times 10^{-1}$ & $-2.62\times 10^{-2}$ & 0.462 \\
        MBH\_3\_3 & 9 & 3 & 0.202046 & -11.920815 & -11.920815 & 0 & $2.94\times 10^{-1}$ & $-3.18\times 10^{-1}$ & $2.08\times 10^{-2}$ & 0.508 \\
        MBH\_3\_3 & 10 & 3 & 0.201743 & -10.797952 & -15.463889 & 0 & $3.38\times 10^{-1}$ & $-2.15\times 10^{-1}$ & $2.41\times 10^{-2}$ & 0.505 \\
        MBH\_3\_3 & 11 & 3 & 0.196205 & -8.774983 & -18.932464 & 0 & $3.08\times 10^{-1}$ & $-1.61\times 10^{-1}$ & $4.26\times 10^{-3}$ & 0.460 \\
        MBH\_3\_3 & 12 & 3 & 0.195407 & -5.852838 & -22.116294 & 0 & $3.47\times 10^{-1}$ & $-1.31\times 10^{-1}$ & $-2.18\times 10^{-2}$ & 0.462 \\
        \hline
    \end{tabular}
    }
\end{table*}

\section{Newtonian N-body simulation}
\label{app:Newtonian}
\begin{figure}
    \centering
    \includegraphics[width=1.0\linewidth]{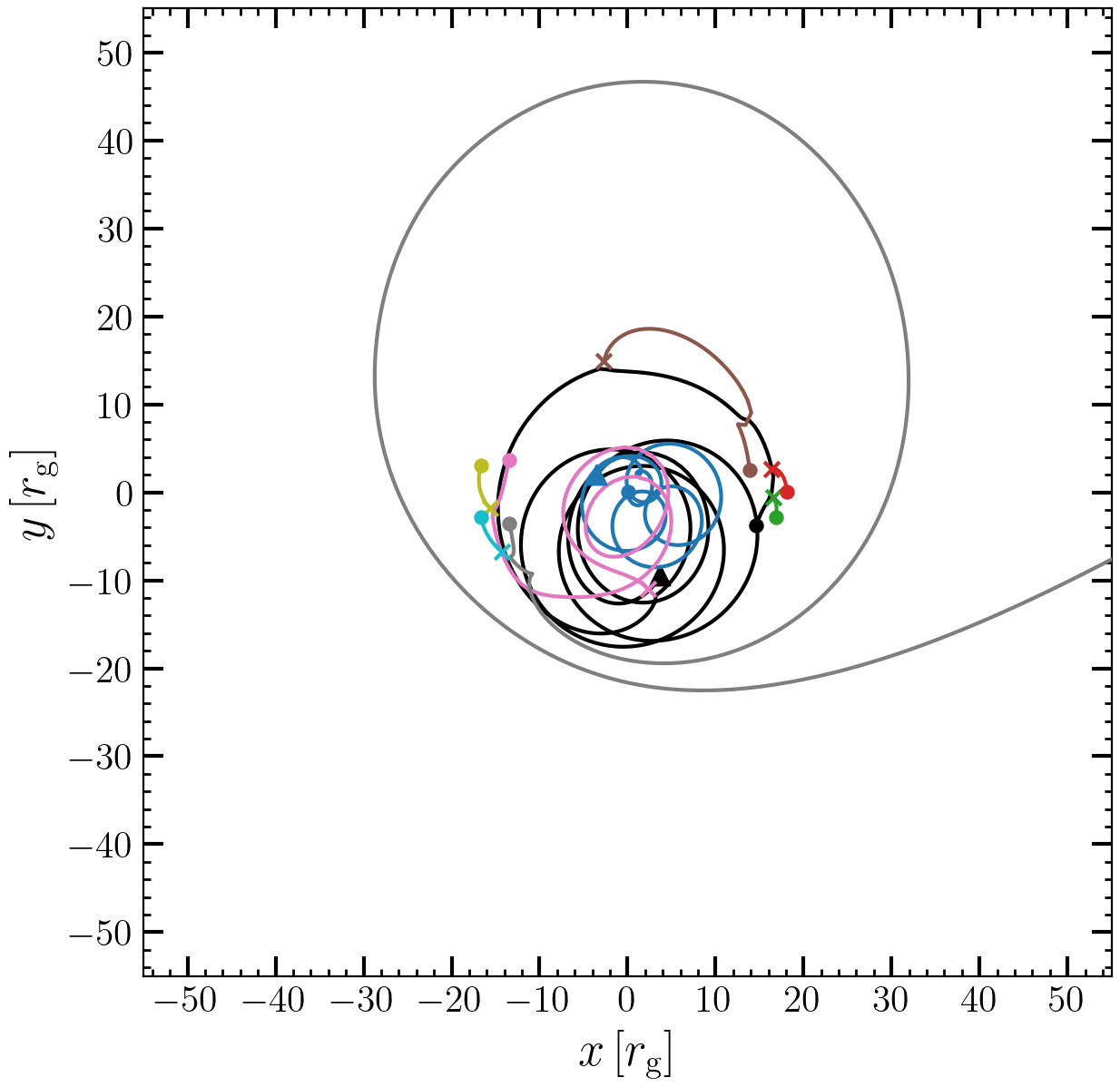}
    \caption{Same as Fig.~\ref{fig:traj_ecc}, but using Newtonian point-particle dynamics.}
    \label{fig:Newtonian}
\end{figure}

For comparison, we have solved the same initial conditions for point masses using a Newtonian N-body integrator scheme \citep{Rein:2011uw}. This includes merging two point masses under conservation of mass and momentum whenever two masses come within their combined Schwarzschild radius. The resulting orbital trajectories are shown in Fig.~\ref{fig:Newtonian}. We can see that the initial focusing and orbital crossing dynamics are almost identical to those in full general relativity (Fig.~\ref{fig:traj_ecc}). However, heading toward merger, the absence of post-Newtonian dynamics substantially alters the resulting picture, including the absence of strong apsidal precession.

\bibliographystyle{aasjournalv7}
\bibliography{inspire,non_inspire}

\end{document}